\documentclass{ws-procs961x669}        

\usepackage[utf8]{inputenc}

\usepackage[cal=cm,scr=dutchcal]{mathalfa}

\newcommand{\rp}{\right)}
\newcommand{\lp}{\left(}

\newcommand{\rcb}{\right\rbrace}
\newcommand{\lcb}{\left\lbrace}

\newcommand{\ra}{\rightarrow}

\newcommand{\pr}{\partial}

\newcommand{\lie}{\mathcal{L}}
\newcommand{\inc}{\lrcorner}

\DeclareMathOperator{\diff}{Diff}

\DeclareMathOperator{\sgn}{sgn}

\DeclareMathOperator{\id}{id}
\DeclareMathOperator{\R}{\mathbb{R}}

\newcommand{\gwg}{\quad \mbox{where} \quad}

\newcommand{\gsotg}{\quad \mbox{so that} \quad}

\def\rn{{\mathrm{n}}}

\def\bh{{\bar{h}}}
\newcommand{\rg}{{\mathrm{g}}}
\newcommand{\U}{\Upsilon}

\def\uk{{\underline{k}}}

\def\um{{\underline{m}}}

\begin{document}
\title{Adiabatic Solutions in General Relativity \\as Null Geodesics on the Space of Boundary Diffeomorphisms}

\author{Emine Şeyma Kutluk$^*$}

\address{Physics Department, Middle East Technical University,\\
Ankara, 06800, Turkey\\
$^*$E-mail: ekutluk@metu.edu.tr\\
www.metu.edu.tr}

\begin{abstract}
We use Weinberg’s trick for adiabatic modes, in a Manton approximation for general relativity on manifolds with spatial boundary. This results in a description of the slow-time dependent solutions as null geodesics on the space of boundary diffeomorphisms, with respect to a metric we prove to be composed solely of the boundary data. We show how the solutions in the bulk space is determined with the constraints of general relativity. 

To give our description a larger perspective, we furthermore identify our resulting Lagrangian as a generalized version of the covariantized Lagrangian for continuum mechanics. We study the cases of 3+1 and 2+1 dimensions and show for the solutions we propose, the Hamiltonian constraint becomes the real homogeneous Monge-Ampere equation in the special case of two spatial dimensions.
\end{abstract}


\bodymatter

\section{Introduction and Summary}

Asymptotic symmetries and soft theorems has been a topic that drew much attention in the last decade. Starting from 2013, Strominger have shown \cite{stro-1,stro-2,stro-3} that soft theorems arise as Ward identities of asymptotic space-time and/or gauge symmetries. Since then many new versions of soft theorems have been suggested \cite{}. \\

Independently of this line of research, in quantum cosmology people have been studying similar Ward identities \cite{cosm-1,cosm-2,cosm-3}. These studies heavily relies on Weinberg's argument for the existence of adiabatic modes on a generic cosmological background \cite{wein-adb}.\\

We use a similar argument for general relativity in a Manton approximation \cite{diet-1,diet-2}. Manton approximation tells one slow time trajectories on the space of static and minimum energy solutions, called the vacua, are good approximate solutions. In our approach we take the vacua to be gauge transformations of a reference metric; then introduce the time dependency in a way that the resulting trajectory is no longer a redundant gauge transformation-as was the case in Weinberg's argument. We study what becomes of the Lagrangian and the equations of motion for these trajectories, and see
\begin{itemlist}[3]
\item Lagrangian solely depends on the quantities on the boundary of spatial slices,
\item constraints of general relativity serves to determine the bulk metric from its boundary values,
\item solutions are geodesics on the space of vacua, which are further required to be null due to the Hamiltonian constraint.
\end{itemlist}

Trajectories on diffeomorphism groups were previously used to describe the motion of ideal fluids \cite{arno} (see also Ref.~\citenum{holm}), where they are described as geodesics with respect to a metric. We covariantize this description \cite{diet-2}, and show that our aforementioned solutions correspond to a specific choice of a generalized metric.\\

As particular examples we study solutions produced from a flat metric on 3-dimensional ball and also general two dimensional enclosed surface. We show how Hamiltonian constraint enforces coupling of two independent towers in the former and how it reduces down to homogeneous-real Monge Ampere equation in the latter case.

\section{A Manton Approximation for General Relativity}
\subsection{Setup}
We start by reviewing the Manton approximation \cite{manton} (see also Refs.~\citenum{stua,kutl-thesis}). Let us have a set of fields $\phi^I(t,x)$, on a spacetime described by the coordinates $(t,x)$.  Let us describe our system with a \textit{natural} action 
\begin{equation}
S= \int  \lp \frac{1}{2} g_{I J} \lp \phi \rp \dot{\phi}^{I} \dot{\phi}^{J}- V \lp \phi \rp \rp dV^{d+1}
\end{equation}
where $dV^{d+1}$, with a slight abuse of notation, denotes the Riemannian volume form for a metric on the $d+1$ dimensional space-time manifold. Let us consider the configuration space, i.e. the space of static fields, called $\mathcal{F}$. 
\begin{figure}
\centering
\includegraphics[width=0.5\linewidth]{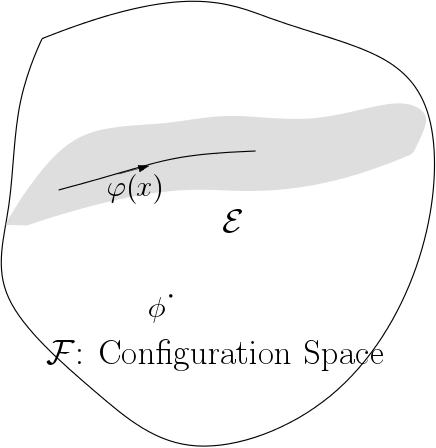} 
\label{fig:minsp}
\caption{Configuration space and the space of minima}
\end{figure}%

Assume that the potential have a continuous set of minima called $\mathcal{E}$; more specifically, in the configuration space let there be a submanifold such that
\begin{equation}
\lp \pr_I V \rp ( \varphi^I ) = 0  \gwg \varphi^I \in \mathcal{E} \ .
\end{equation} 
Note that these static fields are solutions to equations of motion for the type of theory we are considering, since the equation of motion is given by
\begin{equation}
g_{I K} \lp \ddot{\phi}^{I} +  \Gamma^{I}_{M J} \dot{\phi}^{M} \dot{\phi}^{J} \rp = - \pr_{K} V \ .
\end{equation}
Now the metric $g_{IJ}$ will induce a metric on $\mathcal{E}$. The Manton approximation tells us that a \textit{geodesic} on $\mathcal{E}$ with respect to this metric, that is slow in time, is a good approximate solution and the motion off $\mathcal{E}$ can be ignored. \\

Now we want to apply this procedure to general relativity.
For this we need a Lagrangian for general relativity that is of the natural form. This can be achieved by considering a spacetime manifold in the form of $\R \times M$ with the ADM action, and then choosing Gaussian normal coordinates (GNC) where $g_{00}=-1, g_{0i}=0$. Then, renaming the spatial part of the metric as $h_{ij}=g_{ij}$ the action we have is
\begin{equation} \label{eq:Lag}
S=\frac{1}{2} \int  \, \rg_h(\dot{h}_{ij},\dot{h}_{kl}) \, dt - \int \, V[h] \, dt
\end{equation}
where 
\begin{align}
\rg_h(\delta_1 h,\delta_2 h) &= \frac{1}{2}\int_M  \left( h^{ik} h^{jl} -  h^{ij} h^{kl} \right) \delta_1 h_{ij} \delta_2 h_{kl} \, dV^d \ , \label{eq:vac-met}\\
V[h] &= -\int_M  \, R(h) \, dV^d \ .
\end{align} 
Here $dV^d$ is the Riemannian volume form and $R(h)$ Ricci scalar for $h$.$\rg_h$ will be the metric on the configuration space. Note that since $g_{00},g_{0i}$, generally known as lapse and shift, are Lagrange multipliers which we have fixed; constraints of GR 
\begin{align}
  & \nabla^i \left( \dot h_{ij}- h^{kl}\dot h_{kl} h_{ij} \right) =0 \, , \\
  & {R}(h)+\frac{1}{2}h^{ij}h^{kl}\dot h_{i[j}\dot h_{k]l} = 0 \, , 
\end{align} 
needs to be imposed externally from now on. \\

To implement the Manton approximation, we now need to identify the set of static solutions that extremize the theory. Note that the potential is not bounded from below, so there will be no minimal energy. However it will have an extrema:
\begin{equation}
\left. \delta V \right|_{\bar{h}} =0 \ra \quad R_{ij}(\bh)=0 \ .
\end{equation}
So the Ricci flat metrics $\bar{h}$ and their gauge transformations will compose the space of static set of extrema of the action. For our investigation we focus on a single reference metric and its gauge transformations, i.e. we define the static extrema $\mathcal{E}$ to be composed of metrics 
\begin{equation}
\mathscr{h} = \psi \cdot \bar{h} \ ,
\end{equation}
i.e. a Ricci flat metric and its gauge transformations. Remember we have chosen a gauge, thus the remaining gauge transformations are not full set of diffeomorphisms. They turn out to be two sets that can be written as
\begin{align}
\psi_{\chi} &= \left( 0, \chi^i(x) \right) \ , \\
\psi_{\alpha} &= \left( \alpha(x),\partial_j \alpha(x)\int_0^t h^{ij}(t',x)dt' \right) .
\end{align}
First of these are simply the spatial diffeomorphisms, whereas the second type are a form of local boost- they are the transformations that take a choice of ``initial" hypersurface to another one when choosing GNC, see Ref.~\citenum{kutl-thesis} for more details. Local boosts are field-dependent transformations, and thus they do not form a proper algebra- reflecting the fact that constraints of general relativity do not form a proper algebra.\footnote{Commutators of constraints of general relativity gives what is called an algebroid, see  Ref.~\citenum{isham} and Ref.~\citenum{bojowald} for further discussion.} Moreover they are not static as what a Manton description tells a vacuum ought to be. Because of these reasons we restrict our attention to spatial diffeomorphisms and define $\mathcal{E}$ as the space of metrics produced by spatial gauge transformations of the flat metric. \\

So we take $\mathcal{E}$ to be the set of spatial metrics that are -spatial- gauge transformations of a spatial reference metric, i.e.
\begin{equation}
h=\phi^* \bar{h} \ .
\end{equation}
Written in a coordinate basis this reads as
\begin{equation}
h^z_{ij}(x)= \frac{\partial \phi^\uk(x,z)}{\partial x^i} \frac{\partial \phi^\um(x,z)}{\partial x^j} \bar{h}_{\uk \um} 
\end{equation} 
where z parametrizes the gauge transformations. The dynamics is then described as a motion on this space:
\begin{equation}
h^z_{ij}(x,t)= \frac{\partial \phi^\uk(x,z(t))}{\partial x^i} \frac{\partial \phi^\um(x,z(t))}{\partial x^j} \bar{h}_{\uk \um} \ ,
\end{equation} 
or 
\begin{equation}
h(t)=\phi_t^* \bar{h} \ .
\end{equation}
Note here the ``trickiness": Once we introduce the time dependency this way, what we are doing is not a coordinate/gauge transformation in the full spacetime. \\ 

Let us illustrate this with a simple example in 3+1 dimensions. Take reference spatial metric to be the flat metric $\bar{h}_{ij}=\delta_{ij}$. Consider the transformations
\begin{equation}\label{coo-tfm}
\phi^i_t= a(t) x^i \ .
\end{equation}
For these transformations our ansatz will produce the spacetime metric
\begin{equation}
ds^2= -dt^2+ a^2(t) (dx^2+dy^2+dz^2) \ .
\end{equation} 
This illustrates very clearly that our ansatz gives us physically new solutions, for this case the spatially flat FRW spacetime. \footnote{This will not solve vacuum Einstein equations, but this is not relevant for the point made here.}

\subsection{Motion as a Geodesic on the Space of Boundary Diffeomorphisms}

In the last subsection we have proposed solutions to Einstein equations in GNC where
\begin{equation}\label{eq:slns}
h(t)=\phi_t^* \bar{h} \ .
\end{equation} 
Now we would like to check whether these solve the Einstein equations or if they should satisfy further conditions. Note that our free variable is the diffeomorphism $\phi(t,x)$ here. We start by first checking the constraint equations that we have started imposing externally after we have eliminated the Lagrange multipliers. First we observe for our solutions \eref{eq:slns} we have
\begin{equation}\label{delh}
 \dot{h}_{ij} = \nabla_i \xi_{j} + \nabla_j \xi_{i} \quad \gwg \quad \xi^i := \dot{\phi}^{\uk} \frac{\pr x^i}{\pr \phi^{\uk} } \quad \gsotg \quad K_{ij} = \nabla_{(i} \xi_{j)} \ .
\end{equation} 
Here $\nabla$ is the covariant derivative belonging to $h$. Using this and the fact $R_{ij}(h)=R_{ij}(\bar{h})=0$, momentum constraint becomes 
\begin{equation}
\mathcal{P}_j(\xi)=\nabla^i \nabla_{[i} \xi_{j]}=0 \ ,
\end{equation}
whereas Hamiltonian constraint is
\begin{equation}
\mathcal{H}(\xi)=(\nabla_i\xi^i)^2-\nabla_{(i}\xi_{j)} \nabla^i \xi^j=0 \ .
\end{equation}
Furthermore the dynamics will be determined with the reduced Lagrangian, where one plugs \eref{eq:slns} in \eref{eq:Lag}. For our solutions the potential is zero whereas the kinetic term is the length of $\dot{h}$ with respect to the metric
\begin{equation}\label{eq:vac-met-2}
\rg_h(\delta_1 h,\delta_2 h) = \frac{1}{2}\int_M  \left( \nabla_{(i}\xi_{1j)} \nabla^i \xi^j_2 -(\nabla_i\xi^i_1)(\nabla_i\xi^i_2) \right) dV^d \ .
\end{equation}
This is the metric \eref{eq:vac-met} on the configuration space pulled back to the space of vacua. As we have noted, solutions to the dynamical equation of the above action will be geodesics on the space of vacua with respect to this metric. Then we see \textit{Hamiltonian constraint restricts these geodesics to be null.} \\

Now note that each spatial slice is diffeomorphic to the reference slice, so everything can be re-expressed in coordinates where the spatial metric is equal to reference metric i.e. defining
\begin{equation}
\sigma := {\phi_t}_* \xi \,
\end{equation} 
constraint equations and the metric on the vacua space can be rewritten as
\begin{align}
\mathcal{P}_j(\sigma) &=\bar{\nabla}^i \bar{\nabla}_{[i} \sigma_{j]}=0 \ , \label{eq:mom-eq-3}\\
\mathcal{H}(\sigma) &=(\bar{\nabla}_i\sigma^i)^2-\bar{\nabla}_{(i}\sigma_{j)} \bar{\nabla}^i \sigma^j=0 \ , \label{eq:ham-eq-3} \\
\rg_h(\delta_1 h,\delta_2 h) &= \frac{1}{2}\int_M \left( \bar{\nabla}_{(i}\sigma_{1j)} \bar{\nabla}^i \sigma^j_2 -(\bar{\nabla}_i\sigma^i_1)(\bar{\nabla}_i\sigma^i_2) \right) \, dV^d(\bar{h}). \label{eq:vac-met-3}
\end{align} 
We will use this coordinate frame for ease of calculation. Note that coordinate transformation to this frame will be different for each spatial slice.\\

Now we proceed to showing that the metric on the space of vacua, \eref{eq:vac-met-3}, is composed of the boundary data. For this we would like to decompose things into parts tangent and normal to the boundary $\pr M$, so we simply write everything in the basis $\lcb n, e^a \rcb$ where $n$ is the unique normalized vector field orthogonal to $\pr M$ and $e^a$ is a orthogonal basis tangent to $\pr M$.\footnote{We take this to be true also slightly off the $\pr M$.} Now if one uses the momentum constraint \eref{eq:mom-eq-3} within \eref{eq:vac-met-3}, and assume vector field $\sigma$ to be boundary preserving i.e.
\begin{equation}
\left. n_i \sigma^i \right|_{\partial M}=0  \ ,
\end{equation}
one can see that
\begin{equation} \label{eq:vac-met-bnd}
\rg_h(\sigma_{(1)},\sigma_{(2)})=\oint_{\partial M}  \left( \sigma^a_{(1)} D^\perp \sigma_a^{(2)}- \mathscr{K}_{ab}\sigma^a_{(1)}\sigma^b_{(2)} \right) \, dV^{d-1}(\bar{k})
\end{equation} 
where $\mathscr{K}$ is the extrinsic curvature of $\partial M$ in $M$, $\sigma^a$ is the part of $\sigma^i$ that is tangential to $\partial M$, $\bar k$ is the induced metric on the boundary and we have defined a normal derivative
\begin{equation}
D^\perp\chi^a\equiv n^i\partial_i\chi^a+ \Gamma^a_b\chi^b \gwg
\Gamma_a^b = n^i e^b_j \bar\nabla_i e^j_a \ .
\end{equation}
Note that
\begin{itemize}
\item $D^\perp$ is $\bar k$-compatible.
\item For a basis where $\left[n,e_a \right]=0$ one has $\Gamma^a_b=K^a_b$.
\end{itemize}

Looking back at the metric \eref{eq:vac-met-bnd} we see that it is completely composed of the boundary data, except the term with the normal derivative. Remember that $\sigma$ still need to satisfy the constraint equations \eref{eq:mom-eq-3} and \eref{eq:ham-eq-3} in the bulk. If we can show given a vector field on the boundary, there exist a solution to these equations with unique normal derivative $D^\perp \sigma$ on the boundary, we can say we have reduced our action to boundary data. Let us examine these equations and their solutions.\\

We will start with the momentum constraint: Using the theory of differential forms on manifolds with boundary, especially the \textit{Hodge-Morrey-Friedrichs decomposition} following theorem can be proven. \cite{diet-1}
\begin{theorem}
On a homologically trivial manifold with boundary, given a boundary vector field $\zeta^a$, there exists a vector field $\sigma$ such that
\begin{equation}
P(\sigma)=0, \quad  \left. n_i \sigma^i \right|_{\pr M}=0, \quad  \sigma^a= \zeta^a
\end{equation}
and unique up to the gradients of scalars that vanish on the boundary. More specifically $\sigma$ can be written as
\begin{equation}
\sigma=\lp d\alpha+ *d* \beta \rp^\sharp \gwg \rn \beta=0 \ .
\end{equation}
\end{theorem} 
This is true in arbitrary dimensions, and in 3d it takes the form of the Helmholtz theorem. We conjecture that the arbitrariness in $\alpha$ will be fixed by the Hamiltonian constraint, though we were not able to prove this; please see Ref.~\citenum{diet-1} for more details.

However to show that the metric on the space of vacua is solely determined by boundary data one will not need to show this. This is because for the exact part where arbitrariness lies, one can show
\begin{equation}
  \left.D^\perp \partial_a\alpha\right|_{\partial M}=-\mathscr{K}_a^b\partial_b\alpha \, .
\end{equation}

With this, we have shown that the metric on the space of vacua is solely composed of the data on the spatial boundary. \\
Because of this conclusion, we can consider this metric as an inner product on the space of boundary vector fields now: 
\begin{align}
\rg_h \lp \sigma_1,\sigma_2 \rp & = \langle \zeta_{(1)}, \zeta_{(2)}\rangle_h \\ &\equiv \oint_{\partial M}\left( \zeta^a_{(1)} D^\perp \zeta_a^{(2)}-\mathscr{K}_{ab}\zeta^a_{(1)}\zeta^b_{(2)} \right) \, dV^{d-1}(\bar{k}) \ .
\end{align}
\hfill \\
We conclude this section by noting that the space of vacua now can be described as
\begin{equation}
\mathcal{V} \cong  \frac{\mbox{Diff}(\pr M)}{\mbox{biso}(\bar{h})} .
\end{equation}
We quotient by boundary preserving isometries of the reference metric since these will disappear in $\rg_h$, see the original expression \eref{eq:vac-met-2}. This quotient can be performed precisely because of our conjecture that for a given arbitrary boundary diffeomorphism there exists a unique solution to constraint equations implies $\mbox{Diff}(\pr M) \cong \mbox{BDiff}_C(M)$, boundary preserving diffeomorphisms of $M$ that satisfies constraint equations.\\

One can show that $\mathcal{V}$ is a (pseudo-)Riemannian homogeneous space: i.e. the metric $\rg_{\bar{h}}$ we have defined is $\mbox{Diff}(\pr M)$ invariant. This metric will have a mixed signature. Indeed one can see $\langle,\rangle$ is zero for boundary vector fields such that
\begin{equation}
D^\perp \zeta^a - \mathscr{K}_b^a \zeta^b= \mathbb{D} \zeta= 0 \ .
\end{equation}
i.e. if $\zeta \in \ker \mathbb{D}$. One can show that isometries are necessarily in $\ker \mathbb{D}$. If the reverse is not also true the metric will be degenerate on the space of vacua.

\section{Geometric Continuum Mechanics Interpretation}
The idea of describing motion as a trajectory on a diffeomorphism group is well-known in studies of ``ideal continuum motion" \cite{holm}, which are generalizations of Arnold's work on putting motion of an incompressible-ideal fluid in the Lagrangian and Hamiltonian framework \citep{arno}. In the following we summarize a covariantized generalization of ideal continuum motion of Ref.~\citenum{holm}, as presented in Ref.~\citenum{diet-2}. Later on we will see how our formalism of previous section fits within this generalization.\\

Consider a manifold $M$ with a flow $\phi(x,t)$ on it. Derivative with respect to $t$, at $t=0$ of this flow is a vector field on $M$. What about at other times? At other times hardship arises since $\dot{\phi}(x,t) := \partial_t \phi(x,t) $ is the velocity at the point $\phi(x,t)$, not $x$. However one can define
\begin{equation}
\chi(x;t):= \left. \dot{\phi}(y,t) \right|_{y=\phi^{-1}(x;t)} ,
\end{equation}
then $\chi(x;t)$ will be a proper vector field. Now assume there exists a metric $g$ on this manifold $M$, and consider the action
\begin{equation}
S= \frac{1}{2} \int_I \int_M g(\chi, \mathbb{D} \chi) dV_g \, dt
\end{equation}
for some interval $I$ of $t$, where $dV_g$ is the Riemannian volume form for the metric $g$ and $\mathbb{D}: \mathfrak{X}(M) \rightarrow \mathfrak{X}(M) $ is a self-adjoint, positive definite operator. We want to find the equation of motion for this action. Since the operator $\mathbb{D}$ is self-adjoint
\begin{equation}
\delta S =  \int_I \int_M g( \delta \chi, \mathbb{D} \chi) dV_g \, dt .
\end{equation}
Now let us find $\delta \chi$ in terms of $\delta \phi$. We see
\begin{align}
\delta \chi^i(x;t) &=  \delta^i \dot{\phi}(y;t)  - \partial _j \dot{\phi} ( y;t ) \pr_k \lp \phi^{-1} \rp^j \delta \Phi^k(x;t)
\end{align}
where  we set $y=\phi^{-1}(x;t)$ and defined
\begin{equation}
\delta \Phi^i\lp z;t \rp = \delta \phi^j \lp \phi^{-1}(z;t) ;t \rp \ .
\end{equation}
With further modifications we get
\begin{equation}
\delta \chi (x;t)= \pr_t \delta \Phi (x;t) + \lie_\chi \delta \Phi \ ,
\end{equation}
so that
\begin{equation}
\delta S = \int_I \int_M g\lp \pr_t \delta \Phi + \lie_\chi \delta \Phi, \mathbb{D}\chi \rp dV_g dt \ .
\end{equation}
To get variational principle working we should integrate by parts to collect all the operators to one side. For this we note
\begin{align}
\int_M g\lp \lie_\chi \eta, \alpha \rp dV_g &= \int_M \alpha_i \lp \chi^j \nabla_j \eta^i - \eta^j \nabla_j \chi^i \rp dV_g \ ,\\
&= \int_{\pr M} \alpha_i \eta^i \chi^\perp - \int_M \eta^i \lp \nabla_j \alpha_i \chi^j + \alpha_i \nabla_j \chi^j + \alpha_j \nabla_i \chi^j \rp dV_g \ ,
\end{align}
where $\nabla$ is the unique metric compatible and torsion free connection for $g$. We will assume $M$ to have no boundary so that the first term drops. Defining $\lie^\dagger_\chi$ as
\begin{equation}
\int_M g\lp \lie_\chi \eta, \alpha \rp dV_g = \int_M g\lp \eta , \lie^\dagger_\chi \alpha \rp dV_g
\end{equation}
we see that
\begin{equation}
\lie_\chi^\dagger \alpha= - \lp \nabla \cdot \chi \, \alpha + \lie_\chi \alpha + \alpha \inc \lie_\chi g \rp \ ,
\end{equation}
where $\lie_\chi g_{ij}= 2 \nabla_{(i} \chi_{j)}$ and $ \lp \alpha \inc \, \omega \rp_j := \alpha^i \omega_{ij}$. Using this then, we see that the equation of motion becomes
\begin{equation}
\lp \pr_t - \lie_\chi^\dagger \rp \mathbb{D} \chi = 0 \ .
\end{equation}
Writing down explicitly $\lie^\dagger$ and expressing in a basis we get
\begin{equation}
(\partial_t+\zeta^b\nabla_b)\mathbb{D}\zeta^a+(\nabla^{a}\zeta_b+\delta^a_b\nabla_c\zeta^c)\mathbb{D}\zeta^b=0 \ .
\end{equation} 
One can retrieve some classic examples for certain choices of $\mathbb{D}$:
\begin{itemlist}
\item $\mathbb{D}=\id$ gives Euler's equation for the free motion of an incompressible fluid. 
\item $\mathbb{D}=\id-\nabla^2$ gives Camassa-Holm equation for shallow water waves.
\end{itemlist}
As it can be easily identified we get the system of previous subsection for the choice of $\mathbb{D}=D^\perp - \mathscr{K}^a_b$.

\section{Some Results in 3+1 and 2+1 dimensions}
Having described our solutions as null geodesics on the space of boundary diffeomorphisms we now proceed to studying some specific examples.

\subsection{On Round Ball}
We start by considering the case where we take Euclidean 3d ball with round boundary as reference. Let us first look at the momentum constraint.  Expressed in vector calculus form it turns out to be 
\begin{equation}
\vec{\bar \nabla} \times \vec{\bar \nabla} \times \vec{\sigma}=0 \ .
\end{equation}
Solution to this in spherical coordinates will be
\begin{equation}
\vec{\sigma} = a_{\ell m} r^{\ell}  \vec{r} \times \vec{\bar \nabla}Y_{\ell m}(\theta, \varphi) +  \vec{\bar \nabla} \lp f_{\ell m}(r) Y_{\ell m}(\theta, \varphi) \rp \ .
\end{equation}
Note that this satisfies our theorem: $\sigma$ is composed of parts that are exact and coexact with zero normal on the boundary. We conjectured that the exact part should be fixed by the Hamiltonian constraint. However we were not able to solve it even for this simple case.

Plugging in the solution, we get for the metric on the space of vacua 
\begin{equation}
\rg(\sigma^{(1)},\sigma^{(2)})=R^3\sum_{\ell,m} \ell (\ell+1)\left( (\ell-1) a^{(1)}_{\ell m} a^{(2)}_{\ell m} - 2 f^{(1)}_{\ell m}(R) f^{(2)}_{\ell m}(R)\right) \ .
\end{equation}
Notice that even though we do not know $f_{\ell m}(r)$ in the bulk, indeed its bulk value does not appear in the metric, as we have proposed. Moreover Hamiltonian equations forces $\rg(\sigma,\sigma)=0$, thus we see that single towers of $ a_{\ell m} $ or $ f_{\ell m}$ cannot be solutions, there has to be mixing.

On the boundary our solutions have the form
\begin{equation}
\sigma^a = \epsilon^{a b} \partial_b \tau + \bar k^{ab} \partial_b \kappa \ .
\end{equation}
Note that this is basically the Hodge decomposition of a one form on $S^2$, thus our solution space is set of all diffeomorphisms on $S^2$. Except the isotropy group: boundary preserving isometries of $\bar{h}_{ij}=\delta_{ij}$, these will be rotations. For this case the isotropy group will be equal to $\ker \mathbb{D}$ since 
\begin{equation}
\ker \mathbb{D} = \lbrace \sigma^a \, | a_{\ell m} = 0 \, \, \ell > 1 , f_{\ell m}(R)=0 \, \,\forall \ell,m \rbrace = \lbrace \sigma^a = \sum_{m} a_{1 m} \, R \vec{R} \times \vec{\bar \nabla} Y_{1 m} \rbrace  \ .
\end{equation}
Thus our physical vacua space is
\begin{equation}
\mathcal{V}=\mathrm{Diff}(S^2)/\mathrm{SO}(3)
\end{equation}
and our metric on this is non-degenerate.

\subsection{Case of 2 Dimensions}
Since we were not able to solve the Hamiltonian constraint to find solutions in the bulk, we now check if things simplify in $2+1$ dimensions. We will consider a flat reference metric with an arbitrary boundary, and then focus on some specific choices for the boundary. In two spatial bulk dimensions momentum equation gets solved when
\begin{equation}
\vec{\sigma}= c \frac{\pr}{\pr \theta} + \vec{\nabla} \alpha
\end{equation}
where $c$ and $\alpha$ is arbitrary. Hamiltonian equation then becomes
\begin{equation}\label{eq:hrma}
(\pr_x \pr_y \alpha)^2 - \pr_x^2 \alpha \pr_y^2 \alpha = 0 \ ,
\end{equation}
the well-known \textit{homogeneous real Monge-Ampere equation (HRMA)}\cite{polyanin}. Most generalized solution to this is known in a parametric form, together with some family of explicit solutions. However we find that these mostly do not survive the boundary preservingness condition we impose for simple choices of the boundary. \\

Let us first consider some specific examples and start with the case of a round boundary, where $r=R$. The boundary preservingness condition become
\begin{equation}
\left. \pr_r \alpha \right|_{r=R}=0 \ .
\end{equation} 
We find no solution that satisfy this condition. Indeed looking at the metric we see
\begin{equation}
\rg(\chi,\chi)=-\frac{2}{R^2} \int \pr_\theta \alpha(R,\theta) \pr_\theta \alpha(R,\theta) d\theta \ .
\end{equation}
Remember that the metric is derived from the Hamiltonian condition, and thus since it cannot be made to be zero for a non trivial $\alpha$, there exists no non-trivial solution to this problem. (Note that the coexact part is an isometry and will be quotiented out.) \\

Next simplest example is the case of an elliptic boundary. We consider an ellipse that is written as
\begin{equation}
r= \frac{\rho}{2} \frac{\sinh(2 \U)}{\sqrt{\cosh^2\U-\cos^2\varphi}} \ ,
\end{equation}
where $\U$ and $\rho$ are parameters of the ellipse. In this case the boundary preservingness condition is
\begin{equation}
\left. \pr_u \alpha \right|_{\pr M}= - \frac{\rho^2 c}{4} \sin(2v)
\end{equation}
where $u,v$ are elliptic coordinates such that
\begin{align}
x &= \rho \cosh u \cos v \ , \\
y &= \rho \sinh u \sin v \ .
\end{align} 
For this case we find a solution to the HRMA equation
\begin{equation}
\alpha(x,y)= - \frac{\rho^2 c}{2 \cosh \U} \sgn(x) \ y \ F\lp \frac{y}{x} \rp \ ,
\end{equation} 
where $F$ is a hypergeometric function. However this solution has a singularity at the origin. We note here that the HRMA equation is pseudo-elliptic; its elliptic version where the right hand side is positive is more studied and results on existence exists, see e.g. Ref.~\citenum{trudinger}. Because of this we think this type of singularity may not be a surprise in this case. We note that this singularity will be transferred to the resulting metric. We leave further investigation of whether a sensible solution can be obtained from this to the future.

\section{Conclusions}

In this note we summarized the results of Refs.~\citenum{diet-1,diet-2} where using Manton approximation we have proposed some adiabatic solutions of GR produced from diffeomorphisms. We have concluded that an adiabatic solution to GR can be described as a geodesic on the space of vacua 
\begin{equation}
\mathcal{V} \sim \diff(\pr M) / \mbox{biso}(\bar{h}) 
\end{equation}
that are null with respect to the metric
\begin{equation}
\left< \zeta^1, \zeta^2 \right>= \int_{\pr M} \zeta^1_a \mathbb{D} \zeta^a_2 
\end{equation}
that is solely composed of the data on the boundary of spatial slices. We remind that along the way we have used the conjecture that the  Hamiltonian constraint solves for the exact part uniquely. We argued $\mathcal{V}$ is a ``pseudo"-Riemannian homogeneous space with respect to this metric. The metric has mixed but constant signature, and is degenerate if $\ker \mathbb{D} \neq \mbox{biso}(\bar{h})$. We showed our formulation is akin to examples in continuum mechanics in which choice of $\mathbb{D}$ is different. \\

For the specific example of 3 spatial dimensions with round boundary; we explicitly saw our theorems to be holding, we found our vector field only up to exact parts since Hamiltonian equation seems hard to solve. However we saw any non-trivial solution of it has to have both exact and coexact parts. We also have realized $\ker \mathbb{D}=\mbox{biso}(\bar h)$ for this case. In 2 spatial dimensions things got easier in solving the Hamiltonian constraint, since it becomes the Monge-Ampere equation whose solutions are known. For round boundary we found that there exists no non-trivial solution for the vector field in the bulk that is boundary preserving, and for elliptic boundary we were only able to find a solutions that is singular at the origin.\\
\hfill\\
We conclude with a quick overview of the remaining issues. First of all let us list some technical questions still to be worked out:
\begin{arabiclist}[3]
\item Does Hamiltonian equation completely fixes the exact part?
\item Is $\ker \mathbb{D} = \mbox{biso}(\bar{h})$?
\item For 2d case can HRMA be solved for some boundary? 
\end{arabiclist} 
Note that only after we complete solve the bulk vector field, we can construct first $\sigma$ in the bulk of each spatial slice, then we can construct $\xi$ by making the proper transformation for each slice, and only after that one can find $h_{ij}(x,t)$ and the full space-time metric for the adiabatic solutions we propose. \\

Analysis thus far might be suggesting the set of solutions we propose do not survive for vacuum Einstein equations, so next logical step is to perform the analysis for Einstein equations with matter, for example with a cosmological constant. Similarly one can also consider non-compact spatial slices with some fall-off conditions at infinities and study the infrared effects. This type of study would link our work to the studies in asymptotic symmetries and soft theorems and shed a different light to the subject.

\section*{Acknowledgments}
This article summarizes and slightly expands the works in Refs.~\citenum{diet-1,diet-2,kutl-thesis} which was produced in collaboration with Dieter Van den Bleeken and Ali Seraj. I am supported by Research Fund of the Middle East Technical University, Project Number DOSAP-B-105-2021-10763.

\bibliographystyle{ws-procs961x669}
\bibliography{mg-kutluk-biblio}

\end{document}